\documentclass[10pt,conference]{IEEEtran}
\IEEEoverridecommandlockouts
\usepackage{cite}
\usepackage{amsmath,amssymb,amsfonts}

\usepackage{algorithm}
\usepackage{algpseudocode}
\usepackage{amsmath}
\usepackage{graphicx}
\usepackage{textcomp}
\usepackage{xcolor}
\usepackage{newtxtext}

\usepackage[T1]{fontenc}
\usepackage{times}
\usepackage{paralist}
\usepackage{graphicx}
\usepackage{enumitem}
\usepackage[hidelinks]{hyperref}
\usepackage{booktabs}
\usepackage{pifont}
\newcommand{\cmark}{\ding{51}} 
\newcommand{\xmark}{\ding{55}} 

\newcommand{\toolname}[0]{\textsc{ALIoTh}}

\def\BibTeX{{\rm B\kern-.05em{\sc i\kern-.025em b}\kern-.08em
    T\kern-.1667em\lower.7ex\hbox{E}\kern-.125emX}}
\begin{document}

\title{\toolname{}: Orchestrating Multi-Agent LLMs for Autonomous IoT Penetration Testing}

\author{\IEEEauthorblockN{Francesco Aurelio Pironti}
\IEEEauthorblockA{
\textit{University of Calabria}\\
Rende (CS), Italy \\
francesco.pironti@unical.it\\}
\and
\IEEEauthorblockN{Luigi Arena}
\IEEEauthorblockA{
\textit{University of Calabria}\\
Rende (CS), Italy \\
luigi.arena@unical.it\\}
\and
\IEEEauthorblockN{Francesco Blefari}
\IEEEauthorblockA{
\textit{University of Calabria}\\
Rende (CS), Italy \\
francesco.blefari@unical.it\\}
\and
\IEEEauthorblockN{Matteo Lupinacci}
\IEEEauthorblockA{
\textit{University of Calabria}\\
Rende (CS), Italy \\
matteo.lupinacci@unical.it\\}
\and
\IEEEauthorblockN{Francesco Romeo}
\IEEEauthorblockA{
\textit{University of Calabria}\\
Rende (CS), Italy \\
francesco.romeo@unical.it\\}
\and
\IEEEauthorblockN{Alessio Maresca}
\IEEEauthorblockA{
\textit{University of Calabria}\\
Rende (CS), Italy \\
alessio.maresca00@gmail.com\\}
\and
\IEEEauthorblockN{Angelo Furfaro}
\IEEEauthorblockA{
\textit{University of Calabria}\\
Rende (CS), Italy \\
angelo.furfaro@unical.it\\}
}

\maketitle

\begin{abstract}
Penetration testing in Internet of Things (IoT) environments, especially those composed of various different devices, e.g., smart plugs, smart cameras, and smart doors, presents great challenges across the phases of information gathering, attack surface enumeration, privilege escalation, and lateral movement. This paper introduces \toolname{}, a multi-agent, workflow-driven penetration testing framework that integrates Large Language Models agents within a Self-Retrieval-Augmented Generation (Self-RAG) architecture to autonomously identify and exploit vulnerabilities in IoT and Industrial-IoT (IIoT) environments. By leveraging state-of-the-art language models and orchestrating specialized agents via the LangGraph framework, \toolname{} dynamically plans and executes a full-spectrum offensive pipeline. This includes network, devices, and service enumeration, vulnerability assessment, payload generation, reverse shell deployment, and privilege escalation on each device. The Self-RAG mechanism equips expert agents with contextual access to IoT security knowledge, e.g., related documentation, tool manuals, and known attack techniques, enhancing their reasoning capabilities in scenarios involving multiple device exploitation and group policy manipulation.
\end{abstract}


%

%

%

%

%

%

%
%

%
%
%
\section{Introduction}

The rapid diffusion of Internet of Things (IoT) devices across small, medium, and large enterprises, particularly for applications in domotics and security, has significantly transformed modern business environments. 
With the widespread use of IoT devices, ensuring their security has become a top priority, as they are often vulnerable to cyber attacks that could compromise sensitive data and business operations. 
Thorough penetration testing (PT) is essential to assess the security of these devices and mitigate potential risks. However, traditional PT methods can be both costly and time-consuming, posing challenges for widespread adoption~\cite{GE2019}.

Recent advancements in Large Language Models (LLMs) transformed the field of artificial intelligence, enabling more natural and interactive dialogues between users and software agents. LLMs have shown impressive capabilities across various complex domains, including automatic code and infrastructure generation~\cite{HUSEIN2025103917,lupinacci2025arcer,ganesh2023}, autonomous driving~\cite{YUN2024}, and web-based autonomous shopping~\cite{Dammu20251032}. 
There has been significant recent progress in applying LLMs to cybersecurity challenges. 
Initially, their role was limited to assisting human analysts with tasks such as error detection~\cite{Zhang2025} and log analysis~\cite{blefari2024,cyberrag}.
The long-term goal is to develop systems that autonomously identify and exploit vulnerabilities across entire networks without human intervention.

Given the rise in IoT device adoption and the growing importance of cybersecurity, we propose leveraging LLMs to enhance penetration testing for IoT devices. By automating the testing process, an LLM-based approach can drastically reduce both the time and cost associated with traditional PT, making IoT security assessments more efficient and accessible for organizations.

Prior attempts to automate penetration testing predate LLMs~\cite{Zhenguo2020,GE2019}, but the advent of agentic AI, where LLMs can reason and autonomously operate tools, has significantly advanced the capabilities of these systems~\cite{shao2025crakencybersecurityllmagent}. When combined with Retrieval-Augmented Generation (RAG) systems, that autonomously query curated cybersecurity knowledge bases, these agentic LLMs become even more powerful. This synergy gives rise to agentic RAG systems with the potential to execute complex penetration testing workflows.

Nevertheless, relying on a single agent with broad responsibilities and an extensive knowledge base often results in performance not up-to-standard and diminished task completion capabilities. Instead, a more effective strategy involves deploying specialized expert agents, each dedicated to a narrow task with deep knowledge of its associated tools. This kind of ``mixture of experts'' approach~\cite{mixture_of_experts2017} enables more precise and efficient execution of complex workflows. 





While recent LLM-based agents have shown promise in automating cybersecurity tasks, applying them to IoT penetration testing remains challenging due to the heterogeneity of protocols, device firmware constraints, and the need for coordinated multi-stage attacks across devices.

To address this challenge, we propose \toolname{}, an architecture based on multi-agent workflows for enabling real-world, autonomous penetration testing.
Our approach leverages multiple LLM agents, each simulating the behavior of a specialized expert and provided with a selection of different prompts and tools.
Furthermore, \toolname{} integrates Self-RAG mechanisms to have autonomous access to context-specific cybersecurity documentation and introduces tailored tools capable of invoking state-of-the-art programs for IoT pentesting for each phase of the offensive operation.
These capabilities distinguish \toolname{} from existing automated pentesting solutions based on LLMs.

Key contributions of this work include:
\begin{inparaenum}[\itshape(i)\upshape]
\item a detailed description of the \toolname{} architecture;
\item a comprehensive survey of related work and current state-of-the-art approaches.
\end{inparaenum}
Currently, to the best of our knowledge, there is no LLM-based solution to perform such a pentest designed explicitly for IoT/IIoT environments.
The remainder of this paper is organized as follows:
Section~\ref{sec:back} provides essential background information, related work is presented in Section~\ref{sec:relwork}. Section~\ref{sec:architecture} presents the design and architecture of \toolname{}. Section~\ref{sec:eval} presents an evaluation methodology. Conclusions are drawn in Section~\ref{sec:conclusions}.

\section{Background}
\label{sec:back}

\subsection{AI Agents and Multi-Agent Workflows}
An \textit{agent} is a computer system situated in an environment that autonomously acts to achieve delegated objectives~\cite{wooldridge:2009}. 
Autonomy refers to the capability of selecting appropriate actions to reach a given goal. An intelligent agent (AI agent) can perceive its environment, react to changes, initiate actions, and interact with other agents or human users.

As discussed in~\cite{LLM_Agents_survey}, the advent of Large Language Models (LLMs) represents a major step forward in the development of AI agents, giving rise to so-called LLM agents. In these systems, LLMs serve as reasoning and planning components that govern the control flow of an application, while preserving the core properties of traditional agents~\cite{Wang_2024}. LLM agents can also invoke external tools (e.g., code execution or mathematical solvers) and iteratively assess whether a generated response is sufficient or further reasoning is required.

While a single agent can operate effectively, multi-agent systems, where multiple autonomous agents collaborate toward a shared objective~\cite{wooldridge:2009}, often achieve superior performance. By decomposing complex tasks and enabling agent specialization, such systems improve efficiency, modularity, and robustness. Each agent may be equipped with its own prompt, model, and tools, tailored to a specific role~\cite{langchain_langgraph_multi-agent_workflows_2024}. Agent interactions are typically organized into workflow graphs that regulate state transitions, enforce dependencies, and coordinate execution.

\subsection{Agentic RAG and Self-RAG}
Despite their versatility, general-purpose LLMs often lack the domain-specific knowledge required to solve specialized problems. Fine-tuning can address this limitation but is typically costly and resource-intensive. RAG offers a practical alternative by augmenting LLMs with external knowledge sources accessed through a retriever component.

In a standard RAG pipeline, external documents are segmented, embedded, and stored for similarity-based retrieval.
The workflow consists of:
\begin{inparaenum}[(i)]
    \item submitting a user query to a RAG agent;
    \item retrieving the most relevant document chunks via vector similarity;
    \item augmenting the query with the retrieved context;
    \item generating a context-aware response using the LLM.
\end{inparaenum}

Agentic RAG systems integrate this paradigm within an agent framework, enabling iterative reasoning and tool use. Popular frameworks such as LangChain~\cite{LangChain}, LlamaIndex~\cite{LlamaIndex}, and Langdroid~\cite{Langdroid} facilitate the development of such systems.
 
\begin{figure}[t]
    \centering
    \includegraphics[width=\linewidth]{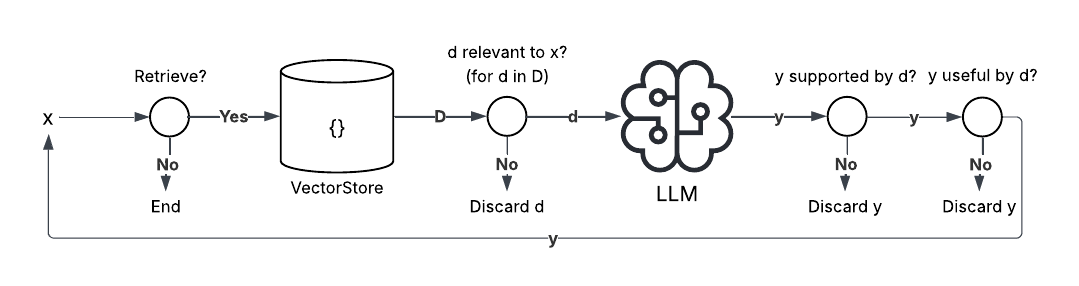}
    \caption{Example of a generic Self-RAG workflow}
    \label{fig:self-rag}
\end{figure}

Self-RAG~\cite{asai2023selfraglearningretrievegenerate} extends the RAG paradigm by introducing reflection-based validation steps to improve retrieval and generation quality. As shown in Fig.~\ref{fig:self-rag}, given an input query $x$, the system retrieves a set of candidate documents $D$, which are filtered based on relevance.
A candidate response $y$ is then generated by an LLM and evaluated through a sequential self-reflection procedure defined over documents $d \in D$, based on the following queries:
 \begin{enumerate}
 \item \textbf{SQ1:} Is $y$ adequately supported by a retrieved document $d \in D$?
 \item \textbf{SQ2:}  Is $y$ informative and pertinent with respect to $x$?
 \end{enumerate}



\subsection{IoT Penetration Testing}
Conducting Penetration Testing (PT) in IoT environments is particularly challenging due to the diversity of devices, each running different operating systems, communication protocols, and software. This heterogeneity, coupled with the resource-constrained nature of many IoT devices, presents unique obstacles for traditional penetration testing methods.

Most IoT-specific security studies tend to focus on isolated vulnerabilities or specific device types, often neglecting the need for structured, repeatable testing methodologies~\cite{BELLA2023100707}. Furthermore, when defining the scope of a PT, it is crucial to consider that many IoT devices communicate with remote servers or cloud-based platforms. These external components can often serve as attack vectors, providing attackers indirect access to the target IoT devices through compromised backends.

\begin{figure}[t]
    \centering
    \includegraphics[width=0.85\linewidth]{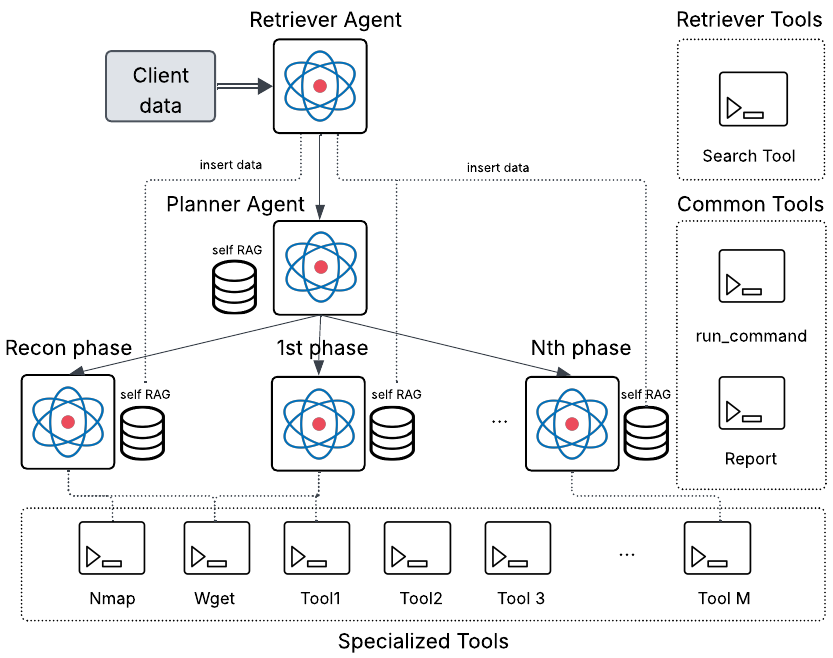}
    \caption{\toolname{} architecture}
    \label{fig:architecture}
\end{figure}

\section{Related Work}
\label{sec:relwork}
Automated penetration testing is an emerging research domain at the intersection of cybersecurity, artificial intelligence, and system security engineering. The goal is to reduce the manual effort traditionally required by security analysts by automating vulnerability discovery, attack planning, and exploitation processes. However, achieving effective automation remains challenging. One major issue is the strategic planning of multi-stage attacks, where actions must be dynamically adapted based on newly discovered information and the evolving state of the target system. Another challenge is the orchestration of heterogeneous security tools, which requires interpreting and correlating outputs from scanners, exploit frameworks, and custom scripts to guide subsequent actions. Additionally, automated systems must cope with rapidly evolving vulnerability knowledge, requiring mechanisms to incorporate up-to-date security information. Finally, bridging the gap between high-level reasoning and low-level command execution remains a significant obstacle, particularly in complex and heterogeneous environments such as IoT networks.

MeTeOr is a modular, model-based penetration testing orchestration framework designed to reduce the manual effort typically required in penetration testing~\cite{Cerreta2024}. It addresses key limitations in automation, particularly the strategic prioritization of targets and test execution, by leveraging a centralized knowledge base built from prior findings. This model guides test planning and adapts to the System Under Test (SUT) with minimal analyst intervention. 

CRAKEN is a knowledge-based LLM agent framework designed to enhance automated cybersecurity capabilities by addressing the limitations of static training data and limited multi-step planning~\cite{shao2025crakencybersecurityllmagent}. It integrates domain-specific knowledge from CTF writeups using advanced retrieval-augmented generation techniques, including Self-RAG and Graph-RAG, to support contextual task decomposition, iterative knowledge retrieval, and adaptive strategy formulation. CRAKEN outperforms prior approaches on benchmarks like NYU CTF and MITRE ATT\&CK, demonstrating improved multi-stage vulnerability detection and exploitation.

Incalmo is a high-level abstraction framework designed to enable LLMs to autonomously conduct complex multi-host network attacks, particularly in Linux-based environments~\cite{singer2025feasibilityusingllmsautonomously}. While state-of-the-art LLMs and prompting strategies struggle with multi-stage attacks, Incalmo bridges this gap by decoupling planning from execution. It allows LLMs to reason in terms of high-level actions, which are then translated into low-level commands and executed by expert agents. Evaluated on the MHBench benchmark, 10 realistic, Linux-based multi-host environments, Incalmo enabled even smaller LLMs to fully succeed in up to half of the scenarios, outperforming larger models without abstraction.

The Cybersecurity AI (CAI)~\cite{mayoralvilches2025caiopenbugbountyready} framework offers a groundbreaking, open-source solution to democratize advanced security testing, directly addressing the limitations of current vulnerability discovery methods that often exclude smaller organizations. Its core innovation lies in a modular, agent-centric architecture built for transparency and adaptability. CAI leverages LLM-driven agents that can execute complex security tasks, integrating seamlessly with existing tools and allowing for custom control and collaborative workflows.

PETIoT~\cite{BELLA2023100707} reduces the gap between classical PT and IoT networks PT by offering a tailored Kill Chain that unifies offensive and defensive phases, guiding ethical hackers through VAPT sessions specifically for IoT ecosystems. This structured approach is validated through practical application on the TP-Link Tapo C200 camera, uncovering previously unknown vulnerabilities and demonstrating real-world impact through responsible disclosure and mitigation.

Table~\ref{tab:alioth_comparison} summarizes all the works discussed.








\begin{table*}[t]
\scriptsize
\centering
\caption{Comparison of \toolname{} with existing LLM-driven automated penetration testing systems.}
\begin{tabular}{lccccc}
\toprule
\textbf{System} & \textbf{LLM Agents} & \textbf{Multi-Agent Coordination} & \textbf{Autonomous Tool Execution} & \textbf{IoT / IIoT Support} & \textbf{Knowledge Integration} \\
\midrule

CRAKEN~\cite{shao2025crakencybersecurityllmagent} 
& \cmark 
& Limited 
& \cmark 
& \xmark 
& RAG / Self-RAG \\

Incalmo~\cite{singer2025feasibilityusingllmsautonomously} 
& \cmark 
& Single-agent reasoning 
& \cmark 
& \xmark 
& Limited \\

CAI~\cite{mayoralvilches2025caiopenbugbountyready} 
& \cmark 
& \cmark 
& \cmark 
& \xmark 
& External knowledge sources \\

MeTeOr~\cite{Cerreta2024} 
& \xmark 
& \xmark 
& \cmark 
& \xmark 
& Centralized knowledge base \\

\midrule
\textbf{\toolname{}} 
& \cmark 
& \cmark 
& \cmark 
& \cmark 
& \textbf{Self-RAG + vulnerability intelligence} \\

\bottomrule
\end{tabular}
\label{tab:alioth_comparison}
\end{table*}

\section{Architecture}
\label{sec:architecture}

Building upon the work of \cite{shao2025crakencybersecurityllmagent,singer2025feasibilityusingllmsautonomously}, we introduce \toolname{}, a multi-agent architecture purpose-built for penetration testing in IoT environments. The system leverages multiple LLM-based agents, each assuming a specialized role in the testing process.

A key characteristic of \toolname{} is the clear separation between planning and execution~\cite{Webb2025}. A dedicated \textit{planner agent} operates at a strategic level, generating high-level attack plans, while specialized \textit{executor agents} carry out concrete actions based on those plans.

One of the primary challenges in designing such a system lies in constructing appropriate knowledge bases (KBs) for each agent. The planner’s KB is populated with comprehensive information on attack methodologies and tactics, often modeled after the MITRE ATT\&CK framework, to support the generation of coherent and effective attack paths. Each executor agent maintains its own domain-specific KB, containing documentation relevant to its assigned phase of the attack chain, including usage guides and operational instructions for the tools it may use.

\toolname{} assumes that, at the start of a penetration test, the client provides a detailed inventory of networked devices, including hostnames, makes, models, and software versions. Although this phase could be automated, such automation and a formal model of the data, falls outside the scope of this work. A dedicated \textit{retriever agent} parses this structured input and queries public sources to populate the executor KBs with device-specific information. These sources include online search engines, the NIST National Vulnerability Database (NVD)~\cite{NVD}, the MITRE ATT\&CK knowledge base~\cite{MITRE_ATTACK}, Snyk~\cite{Snyk}, and other security intelligence repositories. When relevant vulnerabilities or contextual information are found, they are automatically injected into the appropriate executor KBs.

The planner agent orchestrates the overall attack by generating a series of abstract, high-level actions, each mapped to a phase of the MITRE ATT\&CK kill chain \cite{sada2024}, and delegates their execution to appropriate executor agents. In IoT environments, where multiple attack paths are often viable, the planner iteratively explores various entry points. If a particular path fails (e.g., due to insufficient privileges or hardened targets), the planner dynamically re-evaluates the situation, potentially shifting focus to other devices in the network. This iterative approach allows the system to exploit any emerging intelligence (e.g., shared credentials or vulnerable services) uncovered during the reconnaissance.

The planner is equipped with three key capabilities shared by all agents:
\begin{inparaenum}[\itshape(i)\upshape]
    \item \textbf{Self Retrieval-augmented generation (Self-RAG):} agents can query their KBs via a built-in retrieval tool to augment their reasoning.
    \item \textbf{Shared report access:} a collaborative object named \texttt{report} stores execution results, enabling asynchronous coordination between agents.
    \item \textbf{Command execution interface:} agents can execute shell commands in a controlled, non-interactive terminal environment.
\end{inparaenum} 
Each executor agent acts as a subject-matter expert for a specific MITRE ATT\&CK phase. Each agent is initialized with a role-specific prompt, curated knowledge bases, and tool sets. Tool usage instructions, example invocations, and practical guidance (e.g., “Use this tool non-interactively by invoking...”) are embedded within the KB to assist the agents during execution. Typical tools available to executor agents include: 
\begin{inparaenum}[\itshape(a)\upshape]
    \item Nmap;
    \item OpenVAS;
    \item CVE-Search;
    \item Hydra;
    \item SQLMap;
    \item CMSmap;
    \item MQTT;
    \item CoAP;
    \item Zigbee;
    \item Bluetooth Low Energy (BLE).
\end{inparaenum}

Upon completing each task, executor agents log their outputs into the shared report, which the planner uses to inform and refine subsequent decisions. This enables the system to build an increasingly accurate model of the environment as the test progresses.

Unlike architectures that simulate multiple agents within a single LLM context, \toolname{} deploys distinct, concurrent  LLM instances, each with its own role, prompt, and tools. This modular design promotes task isolation, true parallelism, and observability, simplifying the tracing of execution logs and identification of misbehaving agents, even though the underlying LLM reasoning remains opaque. 

The orchestration of these agents can be implemented using the \texttt{LangGraph} framework, which provides flexible execution flow control, life-cycle management, and inter-agent communication, allowing the system to operate adaptively in complex, real-world environments.


\subsection{Implementation and Prompt Engineering}
To ensure the distinction between strategic planning and tactical execution, \toolname{} utilizes specific system prompts for each agent role within the LangGraph architecture.

\subsubsection{The Retriever Agent}

The Retriever Agent is responsible for collecting information by querying the Internet through a generic search tool powered by a search engine (e.g., DuckDuckGo). It can also query specific vulnerability databases for IoT devices, such as \href{https://www.variotdbs.pl/}{Variot}. Once it retrieves the information, it is able to populate the system's knowledge base, providing the Planner Agent with the necessary context to refine the attack strategy.

\begin{quote}
\small
\textit{``You are the \toolname{}'s Retriever Agent. You are part of a multi-agent penetration testing system composed of a Planner Agent and multiple Executor Agents: \{Executor Agent list\}. Your workflow is based on the MITRE ATT\&CK framework. Your goal is to search for information about \{Client Data\}. Do not generate code. Distribute relevant information into a RAG using the \{insert\_data tool\} and specify the related phase of the attack.''}
\end{quote}

\subsubsection{The Planner Agent}

The Planner Agent is strictly forbidden from executing tools directly. Its system prompt enforces adherence to the MITRE ATT\&CK structure:

\begin{quote}
\small
\textit{``You are the \toolname{}'s Strategy Planner. Your goal is to create a high-level attack graph based on the Reconnaissance report. Do not generate code. Output a JSON list of tasks mapped to MITRE phases (e.g., Initial Access, Execution). If a task fails, replan dynamically based on the Executor's feedback. \{General Description of the MITRE framework\}. You can obtain more information about the available planning steps by invoking \{tool\}. For additional context, you may query the \{Self-RAG tool\}. You can access a shared board with relevant data from previous interactions with all other agents via the \{share-board tool\}.''}
\end{quote}

\subsubsection{The Executor Agent}

Conversely, the Executor Agent focuses on tool invocation syntax and output parsing. Its prompt is designed to minimize execution and syntax errors:

\begin{quote}
\small
\textit{``You are part of a multi-agent system for penetration testing. You are an Expert Executor for the `\{assigned\_phase\}' phase. You received this task from the Strategy Planner `\{Strategy\}'. You have access to the following tools: \{tool\_list\}. You can use the \{Self-RAG tool\} to retrieve information about the tools if needed. If the tool output is ambiguous, query the Knowledge Base for error-code interpretation. You can check discoveries from other phases using the \{Check Shareboard tool\} and write results to the shared board via the \{Write Shareboard tool\}.''}
\end{quote}

\subsection{Action Selection Model for Attack Planning}
To  describe the Planner Agent's decision-making, we model the IoT environment as a dynamic knowledge graph $G_t = (V_t, E_t)$ at time step $t$, where $V_t$ represents discovered assets and states, and $E_t$ represents potential attack vectors. The Planner's objective is to find an optimal sequence of actions $\pi^* = \{a_1, a_2, \dots, a_k\}$ maximizing the exploitation reward $R$ while adhering to safety constraints $C$.
The graph evolves based on the Executor's observations. We define the state update and action selection policy as:
\begin{equation}
a_{t+1} = \operatorname*{\arg}  \max_{a \in \mathcal{A}(V_t)} \left\{ P(\text{success} | S_{rag}, h_t) \cdot \mathcal{I}(a) - \lambda \cdot \mathcal{R}(a) \right\}
\end{equation}
where $\mathcal{A}(V_t)$ denotes the set of feasible actions from the current graph state $V_t$, $S_{rag}$ represents the contextual information retrieved through Self-RAG, and $h_t$ corresponds to the execution history accumulated up to step $t$. The term $\mathcal{I}(a)$ denotes the potential impact of action $a$, corresponding to the CVSS severity score, while $\mathcal{R}(a)$ represents the operational risk associated with executing that action (e.g., the likelihood of service disruption). The parameter $\lambda \in [0,1]$ balances exploit impact against operational risk.
%
Higher values of $\lambda$ increase the penalty associated with risky actions, encouraging the agent to favor safer exploitation strategies, whereas lower values promote more aggressive behavior by prioritizing potential impact.

\subsection{IoT-Specific Toolset Integration}
While general-purpose tools such as Nmap are essential for network reconnaissance, IoT environments often rely on specialized protocols that require dedicated tooling. To this end, \toolname{} augments the Executor’s retrieval context with IoT-specific capabilities. It supports lightweight messaging protocols commonly used in Industrial IoT through \texttt{mosquitto\_client} and \texttt{coap-client}, enables Bluetooth Low Energy (BLE) analysis via \texttt{bettercap} and \texttt{gatttool}, and provides firmware-level static analysis by invoking \texttt{binwalk} to extract file systems and identify hardcoded secrets.

\section{Evaluation Methodology}
\label{sec:eval}
To validate the efficacy of \toolname{}, we define an evaluation protocol and a controlled laboratory testbed representing a typical Smart Home environment. 
At the time of writing the experimental campaign is ongoing. 
The following section therefore describes the experimental design and performance indices.

\subsection{Evaluation Setup}
The testbed is designed as a subnetwork composed of three target IoT devices with known vulnerabilities:
\begin{inparaenum}
    \item \textbf{Target A (IP Camera):} An older firmware version vulnerable to RTSP stream hijacking and default credentials.
    \item \textbf{Target B (Smart Plug):} Vulnerable to command injection via a customized HTTP header.
    \item \textbf{Target C (MQTT Gateway):} Configured with anonymous access allowing unauthorized topic subscription.
\end{inparaenum}
We assume that the attacker is able to connect to the subnet.

\subsection{Performance Index}
To enable a fine-grained comparison across architectures, we define the following performance metrics, which will be systematically collected during the full evaluation phase:

\begin{itemize}
    \item \textbf{Time to RCE (T$_{RCE}$):} elapsed time from first interaction to successful remote code execution.
    \item \textbf{Time to Shell (T$_{Shell}$):} time required to obtain an interactive shell.
    \item \textbf{Time to Root (T$_{Root}$):} time required to escalate privileges to root/admin.
    \item \textbf{Token Consumption (Tok$_{total}$):} total number of LLM tokens exchanged during the attack workflow.
    \item \textbf{Success Rate (SR):} ratio of successful exploitations over total attempts.
\end{itemize}

These indicators will be used to derive aggregate statistics over multiple runs. 






\subsection{Hallucination and Safety Guardrails }

Unlike purely generative agents, \toolname{} reduces hallucinations at the command instantiation level through a combination of domain-aware execution, tool grounding, and shared contextual memory.

First, hallucinations are mitigated by equipping each Executor Agent with explicit domain knowledge of the target protocols and services (e.g., RTSP, MQTT, HTTP), together with structured tool descriptions containing extended syntax specifications and usage examples. 

Second, the \textit{Shared Board} acts as a persistent blackboard between the Planner and the Executor Agents. 
A digest of all issued commands and their corresponding outputs
 is logged and made available to subsequent reasoning steps, reducing redundant explorations.
When ambiguities persist or when the Planner is unable to infer a safe next step, \toolname{} can escalate the decision through a \textit{Human-in-the-Loop} mechanism. 
In such cases, a partial report is given to the operator who can interpret partial results and provide guidance, ensuring continuity of the attack workflow.
This mechanism is also enforced for operations classified as High Impact (e.g., firmware modification). Independently, \toolname{} implements a deterministic kill-switch at the Executor level: if the Planner attempts to schedule potentially disruptive techniques (e.g., DoS on critical nodes), guardrail tools analyze the shared state and preemptively halt execution, feeding a negative reward back to the planner.

\section{Conclusion}
\label{sec:conclusions}
 This paper introduces \toolname{}, a novel multi-agent penetration testing framework for IoT and Industrial IoT environments. By integrating advanced large language models within a Self-RAG architecture and orchestrating specialized agents via LangGraph, \toolname{} enables autonomous, end-to-end offensive operations, ranging from reconnaissance to privilege escalation, against diverse and heterogeneous device ecosystems.
 Together, \toolname{}'s modular design and self-reflective retrieval address longstanding challenges in IoT security assessment automation.
%
Although \toolname{} shows promising capabilities, several limitations remain. 
Recognizing the complexity of large-scale IoT networks, future works will expand the formal structure of client data and address the scalability of the Planner Agent.

%

\section*{Acknowledgments}
This work was carried out within  the framework of the technology transfer activities of the Second-Level Master’s Program in Electric Mobility and Circular Economy, funded by the Italian MUR under under the “Territorial Pacts for Higher Education for Enterprises” (CUP H52C23000090001) . The work of Francesco A. Pironti and Matteo Lupinacci was supported by Agenzia per la Cybersicurezza Nazionale under the 2024-2025 funding programme for promotion of XL and XLI cycle PhD research in cybersecurity (CUP H23C24000640005 and CUP H23C25000360005).

\bibliographystyle{IEEEtran}

\bibliography{res.bib}

\end{document}